\documentstyle[12pt]{article}
\makeatletter
\@addtoreset{equation}{section}
\makeatother

\include{figures}
\begin{document}
\begin{titlepage}
\mbox{ }
\rightline{UCT-TP-218/94}
\rightline{October 1994}
\vspace{1cm}
\begin{center}
\begin{Large}
{\bf  Deconfinement and chiral-symmetry restoration in finite
temperature QCD}\footnote{Expanded version of an invited talk at CAM-94,
Cancun, Mexico, September 1994.}
\end{Large}

\vspace{1.5cm}

{\large {\bf C. A. Dominguez}}\footnote{John Simon Guggenheim
Fellow 1994-1995}

Institute of Theoretical Physics and Astrophysics, University of
Cape Town, Rondebosch 7700, South Africa\\

\vspace{.5cm}

\end{center}
\vspace{.5cm}

\begin{abstract}
\noindent
QCD sum rules are based on the Operator Product Expansion of current
correlators, and on QCD-hadron duality. An extension of this program
to finite temperature is discussed. This allows for a study of
deconfinement and chiral-symmetry restoration. In addition, it is
possible to relate certain hadronic matrix elements  to
expectation values of quark and gluon field operators by using thermal
Finite Energy Sum Rules. In this way one can determine the temperature
behaviour of hadron masses and couplings, as well as form factors.
An attempt is made to clarify some misconceptions in the existing
literature on QCD sum rules at finite temperature.
\end{abstract}
\end{titlepage}
\setlength{\baselineskip}{1\baselineskip}
\noindent
\section{Outline}
This report is an expanded version of a talk given at CAM-94, the
joint meeting of the Canadian Association of Physicists, the
American Physical Society, and the Mexican Physical Society.
I discuss here an extension of the QCD sum rule program to
finite temperature. At $T=0$, QCD sum rules offer a very successful
quantum field theory framework to extract information on
hadronic physics from QCD analytically. This great success, unfortunately,
does not translate immediately into the finite temperature
domain, where there are some serious unresolved
problems with Laplace transform sum rules. In spite of this, some
reasonable progress has been made through  lowest moment Finite Energy
Sum Rules (FESR). As with any new field, some degree of confusion
is to be expected as our ideas take shape, and we gain a better
understanding of the subject. It is important, though, not to persist
on wrong notions, nor advocate results which have been shown to be in
contradiction with more fundamental {\it facts}. I shall attempt to
clarify here some misconceptions present in the existing literature
on this subject. The purpose is not to antagonize, but rather to point
out the problems and promote discussion that might help to solve
them. I may group these misconceptions into two categories: (a) conceptual,
and (b) specific results of applications. Among the first is the
statement (without proof) that the notion of QCD-hadron duality, one of
the two pillars of QCD sum rules, should abruptly disappear as soon as
the temperature is turned on (by no matter what small amount). If correct,
this proposition would invalidate a smooth extension of the QCD sum rule
program to $T\neq 0$. I shall argue against this scenario. Second, and
in connection with results of applications, there exist quite a few
determinations of hadron masses based on Laplace transforms
which ignore the underlying problems with these sum rules. Chief among
these problems is the fact that  they are inconsistent with the well
known (and well established) temperature behaviour of the quark
condensate and the gluon condensate. Not surprisingly, the predictions
from this approach are in serious contradiction with the $T$-dependence
of hadron masses obtained in other independent frameworks.\\

The outline of this report is as follows. In Section 2 I introduce briefly
the key ideas behind QCD sum rules at $T=0$, and argue for a smooth
extension of this program to finite temperature. I discuss supportive
evidence for the validity of both the Operator Product Expansion, and
the notion of QCD-hadron duality at $T\neq0$. Section 3 deals with
chiral-symmetry restoration, and in Section 4 I review proposals for
phenomenological order parameters to characterize deconfinement. In
Section 5 I show how these two phase transitions have been related using
a lowest moment FESR-QCD Sum Rule. Section 6 summarizes recent
results for the temperature behaviour of the electromagnetic form factor
of the pion in the space-like region, once again through a lowest moment
FESR. Finally, in Section 7 I concentrate on the problems with Laplace
transform QCD sum rules, as well as with higher moment FESR,
and make critical comments on the existing literature.
Except for a major part of Section 7, the ideas and results presented
here have already been published in the literature, and the reader
may trace them through the list of references.
\section{QCD sum rules}
QCD sum rules at $T=0$  \cite{SVZ} are based on the Operator Product
Expansion (OPE) of current correlators at short distances, suitably
extended to include non-perturbative effects. The latter are parametrized
in terms of a set of vacuum expectation values of the quark and
gluon fields entering the basic QCD Lagrangian. Contact with
the hadronic world of large distances is achieved by invoking the notion
of QCD-hadron duality. This leads to relationships between fundamental
QCD parameters ($\Lambda_{QCD}$, quark masses, vacuum condensates, etc.)
and low energy  parameters (hadron masses, widths, couplings, form factors,
etc.). The values of the vacuum condensates in the OPE cannot be
calculated analytically from first principles, as this would be tantamount
to solving QCD exactly. Instead, they are extracted from certain channels
where experimental information is available, e.g. $e^{+}e^{-}$ annihilation,
and $\tau$ decays \cite{BERTL}. It is also possible, in principle,
to estimate them numerically from lattice QCD.
To be more specific, let us consider the  two-point function
\begin{equation}
\Pi \; (q) = i \int d^{4}x \;\exp (i q x) \;
 <0| T(J(x), J^{\dag}(0)) |0 > \; ,
\end{equation}
where $J(x)$ is a local current built from the quark and/or gluon
fields entering the QCD Lagrangian, and having specific quantum numbers.
In the sequel I concentrate on light quark flavours.
The OPE of $\Pi (q)$ is formally written as
\begin{equation}
\Pi \; (q) = C_{I} < I > + \sum_{r} C_{r} (q) < {\cal{O}}_{r} > \; ,
\end{equation}
where the Wilson coefficients $C_{r} (q)$ depend on the Lorentz indices
and quantum numbers of the external current $J(x)$, and also of the local
gauge-invariant operators ${\cal O}_{r}$ built from the quark and gluon
fields of QCD. The unit
operator $I$ in Eq.(2.2) represents the purely perturbative piece. The
OPE  is assumed valid, even in the presence of non-perturbative effects,
for $q^{2} < 0$ (spacelike), and  $\mid q^{2} \mid \gg$ $\Lambda_{QCD}
^{2}$. In principle, all Wilson coefficients are calculable in
perturbative QCD to any desired order in the strong coupling constant. In
the sequel we shall work at leading (one loop) order for simplicity. The
non-perturbative effects are then buried in the vacuum condensates. Since
these have dimensions, the associated Wilson coefficients fall off as
inverse powers of $Q^{2} = - q^{2}$.
For instance, if the current $J(x)$ in
Eq.(2.1) is identified with the axial-vector current
$ A_{\mu} (x) = : \bar{u}(x) \gamma_{\mu} \gamma_{5} d(x) :$, then with
\begin{equation}
\Pi_{\mu \nu}(q) = - g_{\mu \nu} \Pi_{1} (q^{2}) + q_{\mu} q_{\nu}
\Pi_{0} (q^{2})
\; ,
\end{equation}
one easily finds  \cite{SVZ}
\begin{equation}
4 \pi^{2} \Pi_{0} (q) = - \ln \frac{Q^{2}}{\mu^{2}} +
\frac{C_{4} < O_{4} >}{Q^{4}} + \frac{C_{6} < O_{6} >}{Q^{6}}
+ \cdots \; ,
\end{equation}
where $\mu$ is a renormalization scale, and e.g. the leading vacuum
condensate is given by
\begin{equation}
C_{4} < O_{4} > = \frac{\pi}{3} < \alpha_{s} \;
G^{2} > - 8 \pi^{2} \bar{m}_{q} < \bar{q} q > \; ,
\end{equation}
with $\bar {m}_{q} = (m_{u} + m_{d})/2$, and $<\bar{q} q> = <\bar{u} u>
\simeq <\bar{d} d>$. In Eq.(2.4) a term proportional to $m_{q}^{2}/Q^{2}$
has been neglected. The function $\Pi_{0}(q)$, Eq. (2.4), satisfies
a dispersion relation
\begin{equation}
\Pi_{0} (Q^{2}) = \frac{1}{\pi} \int ^{\infty}_{0} ds \;
\frac{\mbox{Im} \; \Pi_{0}  (s)}{s + Q^{2}} \; ,
\end{equation}
defined in this case up to one subtraction constant, which can be disposed
of by e.g. taking the first derivative with respect to $Q^{2}$ in Eq.(2.6).
The notion of QCD-hadron duality is implemented by calculating the left hand
side of Eq.(2.6) in QCD through the OPE, and parametrizing the spectral
function entering the right hand side in terms of hadronic resonances,
followed by a hadronic continuum modelled by perturbative QCD. In this fashion
one
relates fundamental QCD parameters, such as quark masses, renormalization
scales, vacuum condensates, etc., to hadronic parameters such as particle
masses, widths, couplings, etc.. The convergence of the Hilbert transform,
Eq.(2.6), may be improved by considering other integral kernels. This leads
to other versions of QCD sum rules, such as the Laplace transform,
Finite Energy Sum Rules (FESR), etc..
For instance, the  Laplace transform QCD sum rule for the axial-axial
correlator is \cite{SVZ}
\begin{equation}
{\cal L}\;{ \Pi_{0}(Q^{2})} = \Pi_{0}(M^{2}) =
\int^{\infty}_{0} ds\; \exp(- s/M^{2})\; \frac{1}{\pi}
\; Im \Pi_{0}(s)\; ,
\end{equation}
where $M^{2}$ is the Laplace parameter which plays the role of
$Q^{2}$ as the short distance expansion parameter (for light quark
correlators). Performing the Laplace transform of Eq.(2.4) one finds
\cite{SVZ}
\begin{equation}
\int^{\infty}_{0} ds\; \exp(- s/M^{2})\; \frac{1}{\pi}
\; Im \Pi_{0}(s)\; = \; \frac{M^{2}}{4\pi^{2}}\; (1 +
\frac{C_{4} < O_{4} >}{M^{4}} + \frac{1}{2 !}\;
\frac{C_{6} < O_{6} >}{M^{6}}
+ \cdots )\; .
\end{equation}
One should notice from Eq.(2.8) that in Laplace transform QCD sum rules
{\bf all} condensates are involved. Their numerical importance, though,
is suppressed by inverse powers of $M^{2}$, as well as by factorial
coefficients.\\
Either by using Cauchy theorem, or by expanding Eq.(2.7)
in $M^{2}$, the Laplace transform QCD sum rule Eq.(2.7) is equivalent
to the infinite number of FESR
\begin{equation}
(-)^{N-1} \;C_{2N}<O_{2N}> = 4 \pi^{2} \; \int^{s_{0}}_{0} \;
ds\,\, s ^{N-1}\;\frac{1}{\pi}\; Im \Pi_{0}(s)|_{RES} - \frac{s_{0}^{N}}{N}\; ,
\end{equation}
where N=1,2,..., and $Im \Pi_{0}(s)|_{RES}$ stands for the purely
resonant contribution to the hadronic spectral function (i.e. without
the continuum). For $N=1$, $C_{2} <O_{2}>$ is nothing but a short-hand
notation for the perturbative quark mass insertion ($C_{2} <O_{2}>
\propto m_{q}^{2}$). A practical advantage of these FESR is that now
the vacuum condensates of different dimensionality are effectively
decoupled. This becomes particularly important if the objective is to
extract the values of these condensates from a knowledge of the hadronic
spectral function. In fact, this is the correct procedure followed in
\cite{BERTL} to determine the condensates from data on $e^{+}e^{-}$
annihilation and $\tau$-decays. The decoupling of the different
$C_{N} <O_{N}>$ is also important to study the self-consistency
of the Laplace transform QCD sum rules at $T\neq 0$, as will be discussed
in Section 7.

An extension of this QCD sum rule program to finite temperature
was proposed some time ago in \cite{BS}. This proposal
entails the assumptions that (a) the OPE
continues to be valid, except that now the vacuum condensates will develop
an ({\it a-priori}) unknown temperature dependence, and (b) the notion of
QCD-hadron duality also remains valid. I shall discuss below some
evidence in support of these two assumptions. Notice that
in analogy with the situation at $T=0$,
the thermal behaviour of the vacuum condensates is not calculable
analytically from first
principles. Some model or approximation must be invoked, e.g. the dilute
pion gas approximation, lattice QCD, etc..
The quark, the gluon, and the four-quark condensates at
$T\neq 0$ have thus been estimated in the literature \cite{LEUT}-
\cite{EL}. At finite temperature,
the basic object to be considered is the retarded (advanced) two-point
function after appropriate Gibbs averaging
\begin{equation}
\Pi \; (q,T) = i \int d^{4}x \;\exp (i q x) \; \theta(x_{0})
 << [J(x), J^{\dag}(0)] >> \; ,
\end{equation}
where
\begin{equation}
<< A \cdot B >> = \sum_{n}\; \exp (-E_{n}/T)\; \langle n|A \cdot B|n \rangle
\;/Tr (\exp (-H/T)) \; ,
\end{equation}
and $\mid n>$ is a complete set of eigenstates of the (QCD) Hamiltonian.
The OPE of $\Pi (q,T)$ is now written as
\begin{equation}
\Pi \; (q,T) = C_{I} << I >> + \sum_{r} C_{r} (q) << {\cal{O}}_{r} >> \; ,
\end{equation}
It must be stressed that the states $\mid n>$ entering Eq.(2.11)
can be {\bf any} complete set of states, e.g. hadronic states,
quark-gluon basis, etc..
The hadronic (mostly pion) basis has been advocated in \cite{IO},
while the quark-gluon basis was first used in \cite{BS}. These two
approaches are quite complementary, rather than in conflict, as the
information they provide is somewhat different. The pion basis is well
suited to determine the temperature dependence of vacuum condensates
at low $T$. It does not make use of QCD-hadron duality, and thus has
little relationship to the QCD sum rule program.
On the other hand, use of the quark-gluon
basis allows for a smooth extension of that program to finite temperature.
As it continues to rely on both the OPE and QCD-hadron duality, this
approach provides information on thermal Green functions provided the
temperature dependence of the condensates is known. Since the latter can
be obtained e.g. from using the pion basis in Eq.(2.11), the two choices
of the complete set $\mid n>$ complement each other. However,
it must be kept in mind that the choice of the pion basis, being
a form of the virial expansion, is restricted to low temperatures. This
is not necessarily the case for the quark-gluon basis approach. In fact,
given an expression for the condensates in some framework, accurate
enough for {\bf all} $T$ up to the critical temperature, the QCD sum
rules will provide the $T$-dependence of hadronic matrix elements in
the {\bf same} temperature range.

The validity of the OPE (at any temperature) beyond perturbation theory
cannot be proven from first principles, since one does not know how
to solve QCD exactly. However, at $T=0$ one can solve exactly other field
theories which bear some resemblance to QCD, thus providing evidence
in support of this assumption. For instance, a study of
the (exactly solvable) $O(N)$ sigma model, in the large $N$ limit,
and of the Schwinger model, both in two dimensions, shows that the
short distance approximation to exact Green functions agrees
with the result from the OPE \cite{NOV}. An extension of this analysis
to finite temperature \cite{DLR} shows the same agreement, and thus
supports the assumption of the validity of the OPE in this regime. I
briefly summarize the results of \cite{DLR}.
Let me consider first the $O(N)$ sigma model in 1+1 dimensions
which is characterized by the Lagrangian
\begin{equation}
{\cal {L}} = \frac{1}{2} [\partial_{\mu} \; \sigma^{a} (x)] \;
[\partial_{\mu} \sigma^{a} (x)] \; ,
\end{equation}
where a = 1,...N, and $\sigma^{a} \sigma^{a} = N/f$, with f being the
coupling constant. In the large $N$ limit this model can be solved exactly
(for details see \cite{NOV}), it is known to be asymptotically free, and in
spite of the absence of mass parameters in Eq.(2.13), it exhibits dynamical
mass generation. In addition, in this model there are vacuum condensates,
e.g. to leading order in 1/N : $\langle 0| \alpha |0 \rangle = \sqrt{N} \;
m^{2} \;$ , whereas all other condensates factorize, viz. $\langle 0|
\alpha^{k} |0 \rangle = (\sqrt{N} \; m^{2})^{k} \;$ . The $\alpha$ field is:
$\alpha = f (\partial_{\mu} \sigma ^{a})^{2}/\sqrt{N}$, and we are
interested in the Green function associated with the propagation of quanta
of this $\alpha$ field. We have calculated this Green function at finite
temperature \cite{DLR}. Its imaginary part can be integrated analytically
in closed form and is
\begin{eqnarray*}
\mbox{Im} \; \Gamma (\omega, {\bf q} = 0, T) =
\frac{1}{2 \omega^{2}} \left[ 1 + 3 n_{B} \; (\omega/2T) \right]
\end{eqnarray*}
\begin{equation}
+ \frac{1}{2} \left[ \frac{2}{\sqrt{N}}\; \frac{<< \alpha >>}{\omega^{4}} +
\frac{6}{N} \frac{<< \alpha^{2} >>}{\omega^{6}} + \cdots \right]
\end{equation}
where the first term above corresponds to the perturbative contribution, the
second to the non-perturbative, and $n_{B}$ is the thermal Bose factor.
Equation (2.14) is valid in the time-like region; the space-like region
counterpart vanishes in 2 dimensions. Since the model is exactly solvable,
the thermal behaviour of the vacuum condensates can also be calculated, viz.
\begin{equation}
<< \alpha >> = < \alpha > \left[ 1 + 3 n_{B} (\omega/2T) \right]
\end{equation}
In this case the vacuum condensates contribute to the imaginary part, and as
Eq.(2.14) shows, the thermal dependence of the perturbative piece cannot be
absorbed into the condensates. Hence, no confusion should arise between
perturbative and non-perturbative contributions.\\
Next, I consider the Schwinger model in 1+1 dimensions, with the Lagrangian
\begin{equation}
{\cal {L}} = - \frac{1}{4} F_{\mu \nu} \; F_{\mu \nu} + \bar{\psi} i \;
\gamma_{\mu} \; {\cal {D}}_{\mu} \; \psi
\end{equation}
where ${\cal D}_{\mu} = i \partial_{\mu} + e A_{\mu}$. At T=0 this model has
been solved exactly, and in the framework of the OPE \cite{NOV}. The short
distance expansion of the exact solution coincides with that from the OPE.
Here, we are interested in the two-point functions
\begin{equation}
\Pi_{++} (x) = \langle 0 |T \{j^{+} (x) j^{+} (0) \}|0 \rangle
\end{equation}
\begin{equation}
\Pi_{+-} (x) = \langle 0 |T \{j^{+} (x) j^{-} (0) \}|0 \rangle
\end{equation}
where the scalar currents are: $j^{+} = \bar{\psi}_{R} \; \psi_{L} \;\;$, $%
j^{-} = \bar{\psi}_{L} \; \psi_{R}\;\;$, with $\psi_{L,R} = (1 \pm
\gamma_{5}) \psi/2$. The function $\Pi_{++} (Q)$ vanishes identically in
perturbation theory, and the leading non-perturbative contribution involves
a four-fermion vacuum condensate. We have calculated the thermal behaviour
of these current correlators \cite{DLR} and obtain, e.g. for their imaginary
parts in the time-like region (again, there is no space-like contribution
in 2 dimensions)
\begin{equation}
\mbox{Im} \; \Pi_{++} (\omega, {\bf q} = 0,T) = 0
\end{equation}
\begin{equation}
\mbox{Im} \; \Pi_{+-} (\omega, {\bf q} = 0,T) = \frac{1}{4} \; \left[ 1 -
2n_{F} (\omega/2T) \right]
\end{equation}
Hence, the choice of the fermion basis in the Gibbs average of current
correlators does not imply confusing these fermions with condensates, as
argued in \cite{IO}. As Eqs.(2.19)-(2.20) indicate, (perturbative) fermion
loop terms and (non-perturbative) vacuum condensates develop their own
temperature dependence, which in this particular example happen to be
different.\\

Concerning the notion of QCD-hadron duality, it has been suggested
recently \cite{IO} that it is not applicable at finite temperature.
If correct, this would require a singular dynamical mechanism of a
discontinuous nature in order to invalidate the inter-relationship
between QCD and hadronic parameters effected by duality. No such
mechanism has been proposed in \cite{IO}. That this inter-relationship
would abruptly disappear by raising the temperature from $T=0$ to some
arbitrary small value, say a nano-Kelvin, seems quite unlikely,
especially in the absence of a concrete mechanism to achieve it.
According to the QCD sum rule philosophy, at $T=0$ one
calculates the theoretical left hand side of Eq.(2.6) through the OPE
Eq.(2.2), i.e. one uses quark-gluon degrees of freedom, and duality relates
this QCD part to a weighted average of the hadronic spectral function.
The latter arises from using  hadronic degrees of freedom. At very
low temperatures the hadronic spectrum is expected to
change very little, and the external current will still convert into
quark-antiquark pairs. The temperature dependence of the quark and gluon
condensates is known, and at very low $T$ they also hardly change. Hence,
it is only reasonable to assume that nothing drastic will happen to
duality. There is a sort of {\it temperature inertia} affecting both
QCD and hadronic physics at very low $T$.
At finite temperature, though, there  are some new effects coming into
play, e.g. there are contributions to the QCD and hadronic spectral
functions in the space-like region (as opposed to only the time-like region
at $T=0$), and the heat bath can support condensates with non-trivial
quantum numbers. However, these additional contributions vanish
smoothly as T approaches zero, i.e. they do not introduce any discontinuous
behaviour that would abruptly invalidate the notion of QCD-hadron duality.
At moderate temperatures the hadronic spectrum is expected to suffer some
rearrangement, in pace with changes in the condensates and the
increasing importance of the new analytic structure in the complex energy
plane. By retaining the notion of QCD-hadron duality one is able to relate
quantitatively the temperature dependence of hadronic parameters with
that of QCD parameters, as will be discussed in Sections 5-6.\\
\section{Chiral-Symmetry and its restoration}
I begin by discussing some of the symmetries of the QCD Lagrangian
\begin{equation}
{\cal L}_{QCD} = \bar{\psi} (i \gamma^{\mu} \; D_{\mu} - M)
\; \psi - \frac{1}{4} \; F_{\mu\nu} \; F^{\mu\nu} \; ,
\end{equation}
where
\begin{equation}
F^{\mu\nu} = \partial^{\mu} G^{\nu} - \partial^{\nu} G^{\mu}
+ i \; g_{0} [G^{\mu}, G^{\nu}] \; ,
\end{equation}
with
\begin{equation}
\left\{ \begin{array}{llll}
	  G^{\mu} & = & \frac {1}{2} \; \lambda^{a} \; G_{a}^{\mu} &  \\
		    &   &                                  & (a = 1, \cdots 8) \\
	  D^{\mu} & = & \partial^{\mu} + i g_{0} G^{\mu} &          \\
\end{array}
\right.
\end{equation}

and $M$ the quark mass matrix. For the purposes of this talk I
consider only two quark flavours (up and down). Among the various
symmetries of ${\cal{L}_{QCD}}$ one finds a (global) $SU(2)_{V}$ and
an $SU(2)_{R} \otimes SU(2)_{L}$ symmetry. These Lagrangian symmetries are
explicitly broken by the quark masses. In fact, the vector (I-spin) Noether
current $V_{\mu}^{i} = \bar{\psi} \gamma_{\mu} \; \tau^{i} \psi$, and
the axial-vector current $A_{\mu}^{i} = \bar{\psi} \; \gamma_{\mu} \;
\gamma_{5} \; \tau^{i} \psi$ have divergences
\begin{equation}
\partial^{\mu} V_{\mu} = i (m_{d} - m_{u}) \bar{d} u \; ,
\end{equation}
\begin{equation}
\partial^{\mu} A_{\mu} = i (m_{d} + m_{u}) \bar{d} \gamma_{5} u \; .
\end{equation}
In the limit $m_{u} = m_{d} = 0, SU(2)_{V}$,
and $SU(2)_{R} \otimes SU(2)_{L}$
become exact Lagrangian symmetries. However, this is not what is
usually meant by chiral-symmetry restoration, which refers to the
symmetry of the vacuum. Given a Lagrangian symmetry one must
investigate how it is realized in the states, starting with the
vacuum. According to whether the Noether charges $Q^{i} = \int \;
d^{3}x \; J_{0}^{i}(\vec{x},t)$ annihilate the vacuum or not, one has a
Wigner-Weyl or a Nambu-Goldstone realization of the Lagrangian
symmetry. In the former case particles are classified according to
the irreducible representations of the symmetry group, as in e.g.
$SU(2)_{V}$. The Nambu-Goldstone realization (spontaneous symmetry
breaking) corresponds to a {\it {hidden}} symmetry, as the vacuum
does not share the symmetry of the Lagrangian. This is the case for
$SU(2)_{R} \otimes SU(2)_{L}$ since e.g. there are no (quasi) degenerate
parity doublets in the particle spectrum. No particle classification
is possible in this phase as the vacuum is {\it {contaminated}} by an
arbitrary number of massless (Nambu-Goldstone) bosons carrying non-
trivial quantum numbers. In the case of $SU(2)_{R} \otimes SU(2)_{L}$
the three emerging Nambu-Goldstone bosons are readily identified with
the pion $(\pi^{\pm}, \pi^{0})$, which decays to the (hadronic)
vacuum through the axial-vector current, i.e.
\begin{equation}
\langle 0 | A_{\mu}^{i}(0)|\pi^{j}(p) \rangle =
i \; f_{\pi} \; p_{\mu} \; \delta^{ij} \; ,
\end{equation}
with $f_{\pi} = 93.2$ MeV. In the limit $m_{u} = m_{d} = 0$ the
axial-vector current is strictly conserved and, hence,
$f_{\pi} \; \mu_{\pi}^{2} = 0$. In the Nambu-Goldstone phase
\begin{equation}
\mu_{\pi}^{2} \propto (m_{u} + m_{d}) \rightarrow 0 \;  ,
\end{equation}
\begin{equation}
f_{\pi}^{2} \propto \langle 0 | \bar{u} u + \bar{d} d | 0 \rangle
\neq 0 \;  ,
\end{equation}
with the proportionality constants such that $2f_{\pi}^{2}
\mu_{\pi}^{2} = (m_{u} + m_{d}) \langle 0|\bar{u} u + \bar{d} d|0
\rangle$. A phase transition from a Nambu-Goldstone to a Wigner-Weyl mode
({\bf {chiral-symmetry restoration}}) is characterized by the
vanishing of the order parameter $f_{\pi}$,or alternatively
$\langle 0 | \bar{q} q |0 \rangle$. Clearly, this can only happen at
finite temperature, and if the phase transition does take place this
should happen regardless of whether the quark masses are zero or not.
The numerical value of the critical temperature, though, is
expected to depend on this fact.\\

The temperature behaviour of $f_{\pi}(T)$ at low $T$ $(T \ll
\mu_{\pi})$ has been investigated in chiral perturbation theory with
the result \cite{LEUT}
\begin{equation}
f_{\pi}(T) = f_{\pi}(0) \left[ 1 - \frac {T^{2}}{8f_{\pi}^{2}(0)}
+ {\cal{O}}(T^{4}) \right] \; ,
\end{equation}
in the chiral limit for three flavours. Chiral-symmetry breaking
corrections to Eq.(3.9) have been also calculated \cite{LEUT}, together with
corrections due to massive states \cite{GL}. The low temperature expansion
of $\langle \bar{q} q \rangle_{T}$ has been carried out up to order
${\cal O}(T^{6})$ in \cite{GL}. As the authors
of \cite{LEUT},\cite{GL} have pointed out, this low temperature
expansion should not be extrapolated to the critical temperature, as
it is not valid there. For instance, if the phase transition is of
second order then the critical exponent should be $\frac{1}{2}$,
rather than 2 as one would naively obtain from Eq.(3.9). If one is
interested in the behaviour of $f_{\pi}(T)$ near $T = T_{c}$, other
methods should be used, e.g. the composite operator formalism of \cite{BAR},
which reproduces Eq.(3.9) at low $T$, but gives instead $f_{\pi}(T)
\propto (1 - T/T_{c})^{\frac{1}{2}}$ as $T \rightarrow T_{c}$. This
feature will become particularly important in Section 5.\\
\section{Quark Deconfinement}
At zero temperature the shape of a typical hadronic spectral function
consists of some delta functions plus resonances with increasing
widths, followed by a smooth continuum starting at some threshold
energy $E_{0}$. It is known from fits to actual data that for $E
\tilde{>} E_{0}$ the hadronic spectral functions are well
approximated by perturbative QCD; $E_{0}$ is thus called the
asymptotic freedom (A.F.) threshold. This picture is well supported
by all exisiting experimental data.
Under the assumption that quark deconfinement does take place at some
critical temperature $T_{d}$, one would expect that by increasing $T$
from $T=0$ the resonance peaks in the spectral function should become
broader. At $T = T_{d}$ the resonance widths would then become
infinite, signalling quark deconfinement. This resonance melting with
increasing temperature would be accompanied by a shift of the
asymptotic freedom threshold $s_{0}$ towards threshold. In this
picture the resonance width, and/or the asymptotic freedom threshold
provide a suitable order parameter.\\

In the complex s-plane, bound
states (e.g. the pion) correspond to poles of the S-matrix lying on
the real axis, and resonances correspond to poles located in the
second Riemann sheet, their distance to the real axis being measured
by the width $\Gamma_{R}$. At finite temperature the form of the
Green function will be
\begin{equation}
G(E,T) \propto \frac {1}
{E - M_{R}(T) + \frac {i}{2} \Gamma_{R} (T)} \; .
\end{equation}
In the picture described above, an increase in the temperature will
shift {\bf all} poles farther away from the real axis as
$\Gamma_{R}(T)$ increases. While the mass $M_{R}$ may depend on $T$,
it is not a relevant order parameter. With the second Riemann sheet
poles infinitely far from the real axis, and the A.F. threshold
correspondingly close to the origin, the spectral function is then
described entirely by the QCD continuum extrapolated down to the
kinematic threshold. One may interpret such a phase as a deconfined
phase where all hadrons originally contributing to the spectral
function have melted into quarks or quark-antiquark pairs. As an
example, let us consider the following ansatz for the rho-meson
width
\begin{equation}
\Gamma_{\rho}(T) = \frac {\Gamma_{\rho}(0)} {(1 - T/T_{d})^{\alpha}} \; ,
\end{equation}
where $T_{d}$ is the critical temperature for deconfinement, and
$\alpha$ a positive number. The behaviour of the hadronic spectral
function in the vector channel, with $M_{\rho} (T) = M_{\rho}(T=0)$,
$\Gamma_{\rho}$ as in Eq.(4.2) with $\alpha =0.5$, is illustrated
in Fig. 1 for two values of the temperature: $T=0$ (solid curve),
and $T=T_{d}/2$ (dashed curve).


In the case of stable hadrons, e.g. the pion or the nucleon, for which
$\Gamma(T=0)=0$, we would expect an imaginary part in their propagators
to develop for $T\neq 0$. In this case, these hadronic thermal widths
should be interpreted as damping coefficients of wave packets propagating
through a dispersive medium (heat bath). Their growth with increasing
temperature implies that, with hadrons corresponding to excitations
in this medium, these excitations become less and less important. Combined
with the notion of QCD-hadron duality, this may be interpreted as the melting
of hadrons into their constituents, thus signalling a deconfinement phase
transition.\\

I wish to stress that the relevant order parameter for deconfinement
should be the width and not the mass. Claims have been made
occasionally in the literature that the mass should vanish at the
critical temperature. While this may happen in some cases, it is not
a necessary condition for deconfinement. In fact,
let us consider a stable hadron, i.e. one with zero
width at zero temperature, such as the pion or the nucleon. If with
increasing temperature the mass goes to zero, and nothing happens to
the width, then at the critical temperature one would still see a
peak in the hadronic spectral function at zero energy. The hadron has
not disappeared from the spectrum, and hence has not melted! The only
way a particle can melt is by having a temperature dependent width
such that $\Gamma_{R} \rightarrow \infty$ as $T \rightarrow T_{c}$.
What happens to the mass (defined as the position of the pole on the
real axis) is irrelevant to this argument. Once the resonance
becomes infinitely broad, the spectral function becomes smooth and
should be well approximated by the quark degrees of freedom, i.e. by
perturbative QCD.
Since this proposal was first made  in \cite{DL1} independent supportive
theoretical evidence has become available. In \cite{LS} it has been
shown in the framework of the virial expansion that: (a) at low
temperatures ($T < 50$ MeV) and in the chiral limit ($\mu_{\pi} = 0$)
the mass of the nucleon $M_{N}(T) \simeq M_{N}(0)$, and its width
$\Gamma_{N}(T) \simeq \Gamma_{N}(0) \equiv 0$; (b) At higher
temperatures, and now away from the chiral limit,
both $M_{N}(T)$ and $\mu_{\pi}(T)$ {\bf increase}
slightly (by 4\% and 1\%, respectively, up to $T \tilde{<} 160$ MeV),
while $\Gamma_{N}(T)$ and $\Gamma_{\pi}(T)$ increase substantially,
e.g. at $T = 160$ MeV the ratio of width to mass (imaginary to real
part of the propagator) is about 43\% for the pion and 20\% for the
nucleon. Since the pion and nucleon widths are strictly zero at $T =
0$, this is quite a dramatic effect. Additional evidence for the
approximate constancy of $M_{N}$ and $\mu_{\pi}$ follows from the
sigma model \cite{SIGMA}, which also gives increases over $M_{N}(0)$ and
$\mu_{\pi}(0)$ at the level of a few percent. Further support comes
from a recent calculation of $\mu_{\pi}(T)$ in the composite operator
formalism \cite{BAR2}, showing approximate constancy of the pion
mass over a wide range of temperatures, with a tendency to {\bf
increase} near the critical temperature. QCD sum rules give a similar
T-dependence of the nucleon mass \cite{DL3}. The  temperature behaviour
of the pion and nucleon widths have also been calculated recently
in the framework of the sigma model \cite{DL4}. The results are in
qualitative agreement with \cite{LS}, and are shown in Figs. 2-3
for three different values of the $\sigma$-meson mass:
$M_{\sigma}$ = 400 MeV (a), 600 MeV (b), 800 MeV (c). Concerning the
rho-meson mass, it has been shown \cite{DLX} that to first order in
the virial expansion, unitarity of the $\pi-\pi$ scattering amplitude
requires  that if $\Gamma_{\rho}(T)$ increases with $T$, then $M_{\rho}(T)$
must also {\bf increase} with $T$. I shall come back to this point
in Section 7.
\section{Relationship between deconfinement and chiral-symmetry restoration}
Let us consider the retarded two-point function  Eq.(2.10)
involving the axial-vector current
$A_{\mu}(x) = : \bar{u}(x) \gamma_{\mu} \gamma_{5} d(x):$.
As discussed in Section 2, the QCD sum rule program consists in
calculating the two-point function in the deep euclidean region
in perturbative QCD, adding non-perturbative effects parametrized
in terms of vacuum expectation values of the quark and gluon fields
appearing in the QCD Lagrangian, and relating the result to the hadronic
spectral function by means of a dispersion relation.

In order to find the perturbative QCD behaviour of Eq.(2.10) we
assume a dilute quark gas at temperature $T$ with zero chemical potential.
The virtual quanta associated to the local current $A_{\mu}(x)$ will
convert into $q \bar{q}$ pairs for $q^{2} = \omega^{2} - {\bf q}^{2} >
4m^{2}_{q}$, while for space-like momenta $(\omega^{2} - {\bf q}^{2}) < 0$
there is an additional cut in the complex $\omega$-plane
centered around $\omega = 0$ \cite{BS}.
These two distinct processes contribute to the spectral function as follows

\begin{eqnarray*}
\frac{1}{\pi} \mbox{Im} \Pi_{\mu\nu}^{+}(\omega, {\bf q}) =
\sum_{q} \; \int \; LIPS(\omega, {\bf q}, E_{1},{\bf p}_{1}
E_{2},{\bf p}_{2})
\end{eqnarray*}
\begin{equation}
\times \langle 0|A_{\mu}|q \bar{q} \rangle
\langle q \bar{q}|A_{\nu}|0 \rangle
\times [1 - n_{F}(E_{1}) - n_{F}(E_{2})] \; ,
\end{equation}

for $q^{2} \geq 0$, and

\begin{eqnarray*}
\frac{1}{\pi} \mbox{Im} \Pi_{\mu\nu}^{(-)} (\omega, {\bf q})
= \sum_{q}\sum_{\bar{q}} \; \int \; LIPS(\omega, {\bf q}, E_{1},
{\bf p}_{1}, - E_{2}, - {\bf p}_{2})
\end{eqnarray*}
\begin{equation}
\times \langle q|A_{\mu}|\bar{q} \rangle
\langle \bar{q}|A_{\nu}|q \rangle
\times [n_{F}(E_{1}) - n_{F}(E_{2})] \;
\end{equation}

for $q^{2} < 0$. In the above equations $n_{F}(E)$ is the Fermi-Dirac
distribution function, and the phase space is
\begin{eqnarray*}
LIPS(\omega, {\bf q}, E_{1}, {\bf p}_{1}, E_{2}, {\bf p}_{2}) =
\frac{d^{3}p_{1}}{2E_{1}(2\pi)^{3}} \; \frac{d^{3}p_{2}}
{2E_{2}(2\pi)^{3}}
\end{eqnarray*}
\begin{equation}
\times \delta (\omega - E_{1} - E_{2}) \delta^{(3)}
({\bf q} - {\bf p}_{1} - {\bf p}_{2}) \; .
\end{equation}
It must be stressed that the above two contributions to the spectral function,
arising from different physical mechanisms, are unrelated. The first
contribution ($q^{2} \geq 0$) leads to the usual right- and left-hand cuts
in the complex energy plane, and does not vanish at $T = 0$. The second piece
($q^{2} < 0$) comes from a new cut centered around $\omega = 0$, vanishes at
$T = 0$, and is unrelated to duality.

On the hadronic side, at low temperatures $T < \mu_{K}$, only pions from the
gas will contribute to the spectral function. In addition to
the (time-like) pion pole contribution to $Im \Pi^{(+)}_{\mu\nu}$,
there is a (space-like)  piece in the hadronic spectral function
$Im \Pi^{(-)}_{\mu\nu}$ from the center cut in the complex $\omega$-plane.
The latter involves an integral of
$\langle \pi | A_{\mu} | 2 \pi \rangle \langle 2\pi | A_{\nu} | \pi \rangle$
weighted by Bose factors. However, since $A_{\mu} \propto \varphi
(\varphi^{+} \stackrel{\leftrightarrow}{\partial}_{\mu} \varphi)$
this term appears at order $T^{4}/f^{2}_{\pi}$ in Im$\Pi(s)$. In addition,
the numerical coefficient of this term is very small.
Hence, a parametrization in terms of the
pion pole plus a continuum modelled by perturbative QCD should be a good
approximation to the hadronic spectral function.\\

Using all this information in the FESR Eq.(2.9), suitably extended to
$T\neq 0$, choosing $N=0$, and neglecting $C_{2}<O_{2}>$ (which is
proportional to the quark mass squared) one obtains the
following finite temperature FESR \cite{DL2}
\begin{eqnarray*}
8\pi^{2}f^{2}_{\pi}(T) = \frac {1}{2} \; \int^{s_{0}(T)}_{4m^{2}_{q}} \;
dz^{2}v(z)[3 - v^{2}(z)] \; tanh \; \left( \frac{z}{4T} \right)
\end{eqnarray*}
\begin{equation}
+ \; \int^{\infty}_{4m^{2}_{q}} \; dz^{2}v(z) [3 - v^{2}(z)]n_{F} \;
\left( \frac{z}{2T} \right) \; ,
\end{equation}
where $v(z) = (1 - 4 m^{2}_{q}/z^{2})^{\frac{1}{2}}$. Equation (5.4) is an
eigenvalue equation fixing $s_{0}(T)$ once $f_{\pi}(T)$ is known
independently. The advantage of our choice of Green
function should be evident: at low temperatures ($T < \mu_{K}$) apart
from $f_{\pi}(T)$ there are no other unknown $T$-dependent hadronic
parameters such as masses and widths. For instance, QCD sum rules for
the correlator of two vector currents at $T \neq 0$
\cite{BS},\cite{DN}, involve $M_{\rho}(T)$ and $\Gamma_{\rho}(T)$
which, unlike $f_{\pi}(T)$, are a priori unknown and model-dependent.
I shall come back to this point in Section 7.\\

In the chiral limit the FESR Eq.(5.4) takes the simple form
\begin{equation}
8 \pi^{2} f_{\pi}^{2}(T) - \frac {4 \pi^{2}}{3} T^{2} =
\int^{s_{0}(T)}_{0} \; ds \; tanh \left( \frac{\sqrt{s}}{4T}
\right) \; ,
\end{equation}
where $s_{0}(T = 0) = 8\pi^{2} f^{2}_{\pi}(0)$.
In \cite{DL2} the FESR Eqs.(5.4)-(5.5) were solved using the low temperature
expansion for $f_{\pi}(T)$ from \cite{LEUT}, i.e. Eq.(3.9). The result
is that $s_{0}(T)$ vanishes at a critical temperature $T_{d}$, but
$T_{d} < T_{c}$ ($T_{c}$ being the critical temperature for
chiral-symmetry restoration). However, one should keep in mind
that Eq.(3.9) is not valid in the vicinity of $T_{c}$.
As shown in
\cite{BAR}, if one uses the expression for $f_{\pi}(T)$ from the
composite operator formalism, $s_{0}(T)$ vanishes at practically the same
temperature as $f_{\pi}(T)$, i.e. $T_{d} \simeq T_{c}$. This is illustrated
in Fig.4 (reproduced from \cite{BAR}), where $(s_{0}(T)/s_{0}(0))^{1/2}
\simeq f_{\pi}(T)/f_{\pi}(0)$, except very close to the critical
temperature. Given the uncertainties of the method, this minor
difference can be safely ignored.\\


Independent confirmation of this result may be obtained
by using the first Weinberg sum rule at $T \neq 0$, which in the chiral
limit is given by
\begin{equation}
\frac{1}{\pi} \; \int^{\infty}_{0} \; ds [\mbox{Im} \; \Pi_{V} (s,T) -
\mbox{Im} \; \Pi_{A} (S,T)] = f_{\pi}^{2}(T)
\end{equation}
This sum rule was studied in \cite{DL2} using Eq.(3.9), the result
being essentially the same as with the FESR Eq.(5.5).
However, the authors of \cite{BAR} obtained
$T_{d}/T_{c} \simeq 0.99$ using the more accurate expression for
$f_{\pi}(T)$ from the composite operator formalism.\\

In summary, Finite Energy QCD sum rules at $T \neq 0$ lead to the prediction
that $s_{0}(T)$ vanishes at some critical temperature, which is essentially
the same as that for chiral-symmetry restoration : $T_{d} \simeq T_{c}$,
provided one uses an expression for $f_{\pi}(T)$ valid for all $T$, such
as e.g. the one in \cite{BAR}.
According to the interpretation of $s_{0}(T)$ as a relevant
order parameter for quark deconfinement (see Section 4), one can conclude
that the QCD-FESR provide evidence for the existence of
this phase transition.\\
\section{Pion form factor at finite temperature}
Independent phenomenological evidence for the
deconfinement phase transition in QCD may be obtained e.g. by studying the
thermal behaviour of the electromagnetic form factor of the pion, $F_{\pi}$.
In this case one would expect the size of the pion to increase with
increasing temperature. At the critical temperature the pion radius should
presumably diverge, indicating quark-gluon deconfinement.
In this Section I discuss a recent determination \cite{DL5} of the $T$-
dependence of  $F_{\pi}$ in
the space-like region using a Finite Energy QCD Sum Rule (FESR). The
pion form factor at $T=0$ has been extensively studied in the past with
FESR, as well as with Laplace transform QCD sum rules \cite{fpi0}.
In order to establish some notation, as well as the $T=0$ normalization,
I briefly describe the method at $T=0$ before introducing thermal
corrections.\\

The appropriate object to study is the three-point function
\begin{equation}
\Pi_{\mu \nu \lambda}(p,p',q) = i^{2} \int \; d^{4}x \; d^{4}y \;
e^{i(p'x - qy)}
< 0 |T (A_{\nu}^{\dag}(x) \; V_{\lambda} (y) \; A_{\mu}(0))|0> \,,
\end{equation}

where $A_{\mu}(x) = :\bar{u}(x)\gamma_{\mu}\gamma_{5}d(x):$ is the axial-vector
current, $V_{\lambda}$ is the electromagnetic current, and $q = p'-p$ the
momentum transfer. On general analyticity grounds, the three-point function
Eq.(6.1) satisfies the double dispersion relation
\begin{equation}
\Pi_{\mu \nu \lambda}(p^{2},p'^{2},Q^{2}) = \frac{1}{\pi^{2}} \;
\int_{0}^{\infty} \; ds \; \int_{0}^{\infty} \; ds'
\; \frac{\rho_{\mu \nu \lambda} (s, s', Q^{2})}
	  {(s + p^{2}) (s' + p'^{2})} \,,
\end{equation}
defined up to subtractions, which are disposed of by Laplace improving
the Hilbert transform, or by considering FESR.
The correlator Eq.(6.1) involves quite a few structure functions,
associated with all the Lorentz structures that can be formed with
the available four-momenta. In principle, it should
not matter which particular structure one chooses to project the pion
form factor. Following \cite{fpi0} in choosing the combination
$P_{\mu} P_{\nu} P_{\lambda}$, where $P = p + p'$, the hadronic spectral
function in the chiral-limit reads
\begin{equation}
\rho (s, s', Q^{2})|_{HAD} = \frac{1}{2} \; f_{\pi}^{2} F_{\pi}
(Q^{2}) \delta (s) \delta(s') +
\rho (s, s', Q^{2})|_{QCD} [1 - \theta (s_{0} - s - s')] \,,
\end{equation}
where $s_{0}$ signals the onset of the continuum, $f_{\pi} \simeq 93$ MeV,
and
\begin{equation}
\rho (s, s', Q^{2})|_{QCD} = \frac{3}{16 \pi^{2}} \;
\frac{Q^{4}}{\lambda^{7/2}} \left[ 3 \lambda (x + Q^{2})
(x + 2Q^{2}) - \lambda^{2} - 5 Q^{2} (x + Q^{2})^{3} \right] \,,
\end{equation}
to one-loop order (and in the chiral-limit), with
\begin{equation}
\lambda = y^{2} + Q^{2} (2x +Q^{2}) \,,
\end{equation}
and $x = s + s'$, $y = s - s'$. Since one is interested in writing
the lowest moment FESR for $F_{\pi}$, i.e.
\begin{equation}
F_{\pi}(Q^{2}) = \frac{1}{f_{\pi}^{2}} \; \int_{0}^{s_{0}} \; dx \;
\int_{-x}^{x} \; dy \; \rho(x,y,Q^{2})|_{QCD} \,,
\end{equation}
rather than a Laplace transform QCD sum rule, the non-perturbative
power corrections entering the OPE are of no concern  here
(they contribute to higher moment FESR). The integration region in Eq.(6.6)
has been chosen to be a triangle in the (s,s') plane, with base and height
equal to $s_{0}$. Other choices of the integration region, e.g. a
square region of side $s_{1} \simeq s_{0}/\sqrt{2}$, give similar results.
The solution to the FESR Eq.(6.6) is
\begin{equation}
F_{\pi}(Q^{2}) = \frac{1}{16 \pi^{2}f_{\pi}^{2}}
		     \frac{s_{0}}{(1 + Q^{2}/2 s_{0})^{2}} \,.
\end{equation}
Although not evident from Eq.(6.7), it is important to realize that this
analysis is only valid in the region $Q^{2} \geq 1$ GeV$^{2}$,
where one expects a reasonable convergence of the OPE. This limitation
is of no relevance if one is only interested in the thermal behaviour of
the ratio $F_{\pi}(Q^{2}, T) / F_{\pi}(Q^{2}, 0)$. In any case, as shown
in \cite{fpi0}, Eq.(6.7) provides a reasonable fit to the experimental data
in the region $Q^{2} \simeq 1 - 4 $ GeV$^{2}$, if $s_{0} \simeq$ 1 GeV$
^{2}$.\\

The spectral function Eq.(6.4) at finite temperature was calculated
in \cite{DL5} using the Dolan-Jackiw formalism. After substitution
in Eq.(6.6), the result can be expressed as
\begin{equation}
F_{\pi}(Q^{2},T) = \frac{1}{f_{\pi}^{2}(T)} \;
\int_{0}^{s_{0}(T)} \; dx \; \int_{-x}^{x} \; dy \;
\rho(x,y,Q^{2})|_{QCD} F(x,y, Q^{2}, T) \,,
\end{equation}
with
\begin{equation}
F(x,y,Q^{2},T) = 1 - n_{1} - n_{2} - n_{3} + n_{1} n_{2} + n_{1} n_{3}
+ n_{2} n_{3} \,,
\end{equation}
\begin{equation}
n_{1} = n_{2} \equiv n_{F} \left(| \frac{1}{2T} \sqrt{\frac{x+y}{2}}|
\right) \,,
\end{equation}

\begin{equation}
n_{3} \equiv n_{F} \left( | \frac{Q^{2} + (x-y)/2}
					   {2T \sqrt{\frac{x+y}{2}}} | \right) \,,
\end{equation}
and $n_{F}$ is the Fermi thermal factor.
In the equations above a frame was chosen  such that
$p_{\mu} = (\omega,{\bf 0})$, and $p'_{\mu} = (\omega',{\bf p'})$, in
which case
\begin{equation}
\omega = \sqrt{\frac{x+y}{2}} \hspace{1cm} , \hspace{1cm} \omega' = \frac{x +
Q^{2}}
{2 \sqrt{\frac{x+y}{2}}} \,.
\end{equation}
It has been explicitly checked that the ratio
\begin{equation}
R(T) \equiv \frac {F_{\pi}(Q^{2},T)}{F_{\pi}(Q^{2},0)}
\end{equation}
is essentially insensitive to other choices of frames. For instance, one
may choose $p_{\mu} = (\omega,{\bf p})$, and $p'_{\mu} = (\omega',{\bf - p})$,
which leads to different arguments in the thermal factors, but roughly the
same ratio $R(T)$. The temperature dependence of the continuum threshold,
$s_{0}(T)$, is given by \cite{BAR} (see discussion in Section 5)
\begin{equation}
\sqrt{\frac{s_{0}(T)}{s_{0}(0)}} \simeq \frac{f_{\pi}(T)}{f_{\pi}(0)} \,,
\end{equation}
The ratio Eq.(6.13) is shown in Fig.5 for $Q^{2} = 1$ GeV$^{2}$.
This result for $R(T)$ is in nice agreement with the expectation that as
the temperature increases, $F_{\pi}$ should decrease and eventually
vanish at the critical temperature for deconfinement $T_{d}$.\\

Although the OPE breaks down at small values of $Q^{2}$, one may still
extrapolate the ratio Eq.(6.13) into this region just to study the
qualitative temperature behaviour of the electromagnetic radius ratio
$<r^{2}_{\pi}>_{T}/<r^{2}_{\pi}>_{0}$. Doing this, one finds  that
this ratio increases monotonically with  T, doubling
at $T/T_{d} \simeq 0.8$, and diverging at the critical temperature. This
divergence of $<r^{2}_{\pi}>_{T}$ may
be interpreted as a signal for quark deconfinement. In fact, the
behaviour of $<r^{2}_{\pi}>_{T}$ can be traced back to the temperature
behaviour of the asymptotic freedom threshold $s_{0}(T)$. As $s_{0}(T)$
decreases with increasing $T$, a signature of quark deconfinement, the
root-mean-square radius of the pion increases. This result is in
qualitative agreement with the one obtained in the framework of the
Nambu-Jona Lasinio model \cite{njl}.
\section{Laplace sum rules: do they fail ?}
There are a few papers \cite{fhl}-\cite{h2} devoted to Laplace
transform QCD sum rules at finite temperature. I shall concentrate mostly
in the vector-vector correlator, e.g. the $\rho^{0}$-meson channel. All
of these analyses suffer from the following serious drawbacks.
\begin{center}
\parbox{3em}{(i)}\parbox[t]{10cm}
{A single Laplace sum rule involves three unknowns: the mass and
the width of the $\rho$-meson, $M_{\rho}(T)$ and $\Gamma_{\rho}(T)$,
and the continuum threshold $s_{0}(T)$. I am assuming that the
temperature dependence of the vacuum condensates is a known input.
No trick of magic allows a determination of three unknown quantities
from a single equation. In principle, at least, one could take
derivatives with respect to the Laplace parameter $M^{2}$ in order
to end up with three equations. However, this does not work in practice
because of (ii) below.}

\vspace{.3cm}

\parbox{3em}{(ii)}\parbox[t]{10cm}
{The Laplace sum rule is inconsistent with the known temperature
dependence of the gluon and the quark condensates, as will be
shown below. This inconsistency translates into a breakdown of the
FESR beyond the lowest moment.}

\vspace{.3cm}

\parbox{3em}{(iii)}\parbox[t]{10cm}
{A zero-width approximation has been used. While it is true that
at very low temperatures this approximation does not differ much from a
finite-width parametrization, at intermediate temperatures this is not
the case. Since on physical grounds we expect the width to increase with
increasing $T$ (see Section 4), the zero-width approximation is in
principle self-contradictory.}
\end{center}
Since the Laplace sum rule is equivalent to an infinite number of FESR,
it is simpler to break it up into a series of FESR in order to show (ii)
above. However, this is not strictly necessary; the same conclusion
follows from the Laplace sum rule itself, but the procedure is a bit
more involved. The first three FESR of the type Eq.(2.9) at finite
temperature are (notice the change of normalization, as I am now discussing
the neutral $\rho$-meson channel)
\begin{equation}
\int^{s_{0}(T)}_{0} \;
ds\,\, \frac{1}{\pi}\; Im \Pi_{0}(s,T)|_{RES}
= \frac{1}{8 \pi ^{2}} \int^{s_{0}(T)}_{0} \; ds\,\, tanh(\frac{\sqrt{s}}
{4 T}) \,\, + \frac {T^{2}}{18}\,\, ,
\end{equation}
\begin{equation}
C_{4} <<O_{4}>> =  \int^{s_{0}(T)}_{0} \; ds\,\, s \,\,tanh(\frac{\sqrt{s}}
{4 T}) - 8 \pi^{2} \int^{s_{0}(T)}_{0} \;
ds\,\, s \,\,\frac{1}{\pi}\; Im \Pi_{0}(s,T)|_{RES}\,,
\end{equation}
\begin{equation}
C_{6} <<O_{6}>> = -  \int^{s_{0}(T)}_{0} \; ds\,\, s^{2} \,\,tanh(\frac{
\sqrt{s}}{4 T}) + 8 \pi^{2} \int^{s_{0}(T)}_{0} \;
ds\,\, s^{2} \,\,\frac{1}{\pi}\; Im \Pi_{0}(s,T)|_{RES}\,,
\end{equation}
where $C_{N}<<O_{N}>> = C_{N} <O_{N}> (T)$, the term $C_{2}<<O_{2}>>$ has
been neglected (it is proportional to $m_{q}^{2}$), and the hadronic
resonant spectral function is given by
\begin{equation}
\frac{1}{\pi} Im \Pi_{0}(s,T)|_{RES} =
\frac{1}{48 \pi^{2}} \,\,\,
\frac {M_{\rho}^{4}(T) (1 + \Gamma_{\rho}^{2}(T)/
M_{\rho}^{2}(T))} { (s - M_{\rho}^{2}(T))^{2} +
M_{\rho}^{2}(T) \Gamma_{\rho}^{2}(T)}\,.
\end{equation}

We now have three equations to determine the three unknowns $s_{0}(T)$,
$M_{\rho}(T)$, and $\Gamma_{\rho}(T)$, provided the temperature dependence
of the condensates is used as an input. The gluon condensate in the
chiral limit, and at low $T$, is given by \cite{GL}
\begin{equation}
<<\alpha_{s} G^{2}>>= <\alpha_{s} G^{2}>
-\frac{4 \pi^{2}}{1215} \frac{T^{8}}{f_{\pi}^{4}}
( ln \frac{\Lambda_{p}}{T} -\frac{1}{4} )\, ,
\end{equation}
where $\Lambda_{p} = 275 \pm 65 MeV$. The above result is valid up to
$T \simeq 100-120 MeV$. Beyond these temperatures the contribution of
massive states, as well as chiral symmetry breaking corrections, become
gradually more important. Numerically, though, with $<\alpha_{s} G^{2}>
\simeq 0.1 GeV^{4}$ \cite{BERTL} these corrections are below the
10\% level at $T\simeq 180 MeV$ \cite{LPR}. In view of this I shall
take the gluon condensate to be temperature-independent. Concerning
the dimension $d=6$ four-quark condensate, I assume it is proportional
to $<<\bar{q} q >>^{2}$, and use $<<\bar{q} q >>$ from the analysis
of \cite{BAR}. At dimension $d=4$ and $d=6$ the heat bath can sustain
non-scalar condensates \cite{h2}. Numerically, though, they play a
negligible role in this analysis (I have used the estimates of \cite{h2}).\\

In trying to solve Eqs.(7.1)-(7.3) I found no meaningful global solution for
all three unknowns: $s_{0}(T)$, $M_{\rho}(T)$, $\Gamma_{\rho}(T)$.
As I show next, the reason for the absence of a solution
to all three FESR is that they are inconsistent with the assumed
temperature dependence of the condensates.\\


Let us turn the argument around, and assume a temperature functional form
for $M_{\rho}(T)$ and $\Gamma_{\rho}(T)$, and use the three FESR to
determine $s_{0}(T)$, $C_{4}<<O_{4}>>$, and $C_{6}<<O_{6}>>$. For
definiteness I shall assume that $\Gamma_{\rho}(T)$ is given by Eq.(4.2).
Variations of this functional form do not change the conclusions. For
$M_{\rho}(T)$ I consider two possibilities: (a): $M_{\rho}(T) = M_{\rho}(0)$,
and (b): $M_{\rho}(T)$ as determined using unitarity of the $\pi-\pi$
scattering amplitude, and to lowest order in the virial expansion \cite{DLX}.
The latter analysis leads to a monotonically increasing $M_{\rho}(T)$.
Case (a) has already been discussed in \cite{DL6}; the solutions of the
first two FESR for $s_{0}(T)$ and $C_{4}<<O_{4}>>$, respectively,
are shown in Fig.6. Although $s_{0}(T)$ does decrease
with $T$ as expected, the thermal dependence of the gluon condensate
is definitely wrong. Also, solving the third FESR Eq.(7.3) gives a
quark condensate that decreases with $T$ even faster than the gluon
condensate, again in contradiction with expectations. Next, case (b)
is qualitatively similar to case (a), as may be appreciated from Fig.7.
In this figure, the function $M_{\rho}(T)$ is taken from \cite{DLX}.
Hence, the conclusion from this analysis is that while the first FESR,
Eq.(7.1), does lead to a reasonable result for $s_{0}(T)$, the higher
moment FESR are in contradiction with the known $T$-dependence of the
vacuum condensates. Therefore, the FESR program breaks down beyond the
lowest moment, and so does the Laplace transform. As mentioned before,
this analysis can be carried out directly with Laplace sum rules, without
reference to the FESR, but the conclusions are the same.\\
There is a third possibility, which I find quite attractive (although it
does not solve the above inconsistency problem). This is to consider
the ratio $s_{0}(T)/s_{0}(0)$ as a universal function, i.e. the same for
any channel. Clearly, the value of $s_{0}(0)$ will depend on the particular
channel under consideration, as it is obviously different in the vector
and axial-vector channels, and the nucleon channel, etc.. However, the
ratio  $s_{0}(T)/s_{0}(0)$ may be thought of as a phenomenological
order parameter for deconfinement, in which case it is reasonable to
assume it is channel-independent. In this case, one can use e.g. the
first FESR Eq.(7.1) to predict $M_{\rho}(T)$, and the second FESR
Eq.(7.2) to predict $C_{4}<<O_{4}>>$.
The results of this excercise are shown in Fig.8, where
I have used $s_{0}(T)/s_{0}(0)$ from \cite{BAR} (see Fig.4).
Interestingly enough, the temperature dependence of the $\rho$-meson
mass turns out to be in good agreement with the determination
using unitarity of the $\pi-\pi$ scattering to lowest order in the
virial expansion \cite{DLX}. This supports the validity of the lowest
moment FESR Eq.(7.1), but unfortunately it does not solve the problem
with the higher moment FESR or with the Laplace transform, as
$C_{4}<<O_{4}>>$ comes out wrong.
Whether additional mechanisms could be brought into play in order to
rescue the QCD sum rule program remains an open problem.\\

Concerning the existing analyses of the vector channel based on Laplace
sum rules \cite{fhl}, \cite{ahz}, \cite{h1}, \cite{h2}, they all predict
a {\bf decrease} of the $\rho$-meson mass with increasing $T$. This is
in contradiction with the result of \cite{DLX}, which is quite general.
In the light of the above discussion, it is possible to understand why
these analyses predict the wrong $T$-dependence of $M_{\rho}$. Laplace
sum rules are inconsistent with the known thermal behaviour of the quark
and gluon condensates. There is also a Laplace sum rule determination
of the nucleon mass at finite temperature \cite{az} which gives again
a monotonically decreasing $M_{N}(T)$. This is in contradiction
with a lowest moment FESR analysis \cite{DL3} which predicts an almost
constant nucleon mass. It is also inconsistent with \cite{LS}. The
reason behind the discrepancy is the same as in the vector channel. I
should add that in \cite{az} it is claimed that the nucleon mass should
decrease, since it is proportional to the quark condensate
$<<\bar{q} q>>$. This argument is falacious because the contribution
from the hadronic cut in the complex energy plane centered around the
origin (the so called {\it scattering} term) prevents the nucleon mass
from decreasing \cite{DL3}.

\subsection*{Acknowledgements}
I acknowledge support from the John Simon Guggenheim
Memorial Foundation (USA), and the Foundation for Research Development
(ZA). It is a pleasure to thank the organizers of CAM-94 for a
very enjoyable meeting.


\begin{thebibliography}{99}
\bibitem{DL1}
The proposal that the A.F. threshold be used as a deconfinement order
parameter was made in [3] in connection with the
vector-vector correlator (see also [12]), and in [13] by studying the
axial-axial correlator. The correlation of $s_{0}$ with resonance
broadening was first proposed in [13].
\end{thebibliography}
\end{document}